\documentclass[aps,twocolumn,showpacs]{revtex4}
\usepackage{amssymb}

\usepackage[dvips]{graphicx}
\usepackage{dcolumn}
\usepackage{bm}
\usepackage{epsfig}

\begin{document}

\title{Low energy spin wave excitations in bilayered magnetic manganite La$%
_{2-2x}$Sr$_{1+2x}$Mn$_{2}$O$_{7}$ ($0.30\leq{x}\leq{0.50}$)}
\author{H. Martinho}
\email{hercules@ifi.unicamp.br}
\homepage{http://www.ifi.unicamp.br/gpoms}
\author{C. Rettori}
\affiliation{Instituto de F\'{\i}sica ''Gleb Wataghin'', UNICAMP, 13083-970, Campinas,
SP, Brazil.\\
}
\author{D. L. Huber}
\affiliation{University of Wisconsin, Dept. Phys., Madison, Wisconsin, 53706, USA.\\
}
\author{J. F. Mitchell}
\affiliation{Materials Science Division, Argonne National Lab., Argonne, Illinois, 60439,
USA}
\author{S. B. Oseroff}
\affiliation{San Diego State University, San Diego, California, 92182, USA.\\
}

\begin{abstract}
We have studied the low-temperature behavior of the magnetization and the
specific heat of the bilayered perovskite system $%
La_{2-2x}Sr_{1+2x}Mn_{2}O_{7}$, for $0.30\leq x\leq 0.50$. Our analysis
reveals that below $30$ K the temperature dependence of the magnetization of
the ferromagnetic samples, $x=0.30$, $0.32$, and $0.36$, in a field of $1$T
can be interpreted in terms of the thermal excitations of a two-dimensional
gas of ferromagnetic magnons. The specific heat in zero field for these
samples as for the $x=0.50$ antiferromagnetic one, is linear with
temperature between the range of $1.8$K $\leq $ T $\leq $ $10$K. This
behavior can be also explained by the magnon gas model. By comparing
specific heat measurements in zero field with those taken in a field of $9$T
we are able to extract the lattice and electronic contributions and
determine the in-plane exchange interactions. That are found to be in
reasonable agreement with the values inferred from the analysis of the
magnetization data and also with the values reported by inelastic neutron
scattering studies. In addition, we found that the electronic density of
states obtained for the $x=0.50$ sample is in agreement with previous band
structure calculations.
\end{abstract}

\maketitle

\section{Introduction}

Manganese oxides with perovskite structure have been subject of intense
study in recent years. This is in large part motivated by two main factors:
their potential technological application, related to the Colossal
Magnetoresistance (CMR) effect displayed by many of these oxides, and the
large variety of magnetic/electronic phenomena such as double-exchange,
superexchange, Jahn-Teller effect, charge/orbital ordering related to the
CMR.\cite{ManganiteRev,BilayerRev}

A common way to represent the manganese perovskite oxides is through the
Ruddlesden-Popper series (R,A)$_{n+1}$Mn$_{n}$O$_{3n+1}$ (R = rare-earth; A
= alkaline metal). The basic structure of the series consists of alternate
stacking of rock-salt type block layers (R,A)$_{2}$O$_{2}$ and $n-$MnO$_{2}$
sheets along the $c-$axis, where $n$ represents the number of adjacent MnO$%
_{2}$ sheets, a quantity closely related to the effective
dimensionality of the system. So, the $n=\infty $ series
represents the prototype three-dimensional (3D) CMR material
(R,A)MnO$_{3}$, the subject of many
investigations lately; $n=1$ corresponds to compounds such as La$_{2}$CuO$%
_{4}$, the metal oxide that originates the high-$T_{C}$ superconducting
cuprates family; $n=2$ represents the bilayered magnetic manganite system
(R,A)$_{3}$Mn$_{2}$O$_{7}$.\cite{BilayersDisc} This last system shows
two-dimensional (2D) magnetic and electronic properties associated with a
very large magnetoresistance (MR) $\rho (0)/\rho (H)\sim 20,000$\% at $%
T_{C}\sim 130$K and $H=7$T for
La$_{1.2}$Sr$_{1.8}$Mn$_{2}$O$_{7}$.\cite {BilayersDisc} For
comparison, in the 3D Sr-doped La$_{0.85}$Sr$_{0.15}$MnO$_{3}$ the
MR reaches $\sim 110$\% at $H=15$T around $T_{C}$.\cite{3DCMR}

\begin{figure}[tbh]
\includegraphics[height=8cm]{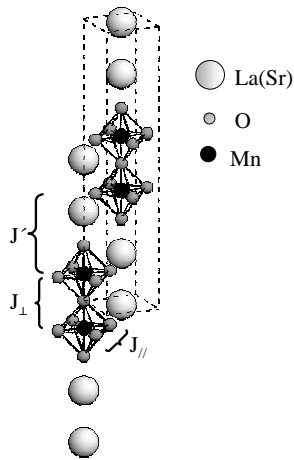}
\caption{Structure of the hole doped bilayered system La$_{2-2x}$Sr$_{1+2x}$%
Mn$_{2}$O$_{7}$. Dashed lines represents the unit cell of the bilayer where $%
J_{\parallel}$, $J_{\perp}$ and $J^{\prime}$ are the exchange constants
described in the text.}
\label{estrutura}
\end{figure}

The structure of the $n=2$ bilayer system is shown in figure \ref{estrutura}%
. It consists of pairs of corner shared MnO$_{6}$ octahedron stacked along
the $c-$axis and isolated by La(Sr) atoms. The dashed lines enclose the unit
cell of the bilayer system. Also, indicated in the figure are the intra-MnO$%
_{2}$ plane, $J_{\parallel}$, inter-MnO$_{2}$ planes, $J_{\perp}$, and
inter-bilayers, $J^{\prime}$, exchange constants.

By changing $x$, the hole-doped bilayers La$_{2-2x}$Sr$_{1+2x}$Mn$_{2}$O$%
_{7} $ present a wide variety of electron transport behaviour and magnetic
structures. For the range $0.30\leq {x}\leq {0.50}$ it is found at low
temperatures, $T<10$K: (i) metallic character and ferromagnetic (FM) order
along the $c-$axis for $x=0.30$;\cite{BilayerMagEstr} (ii) metallic
character and FM order in the MnO$_{2}$ sheets ($ab-$plane) for $x=0.32$; %
\cite{BilayerMagEstr} (iii) metallic character and FM order in the $ab-$%
plane, with a weak antiferromagnetic (WAFM) component along the $c$-axis for
$0.32<x\leq {0.48}$;\cite{BilayerMagEstr} (iv) metallic A-type AFM with a
small fraction ($<18$\%) of a secondary CE-type AFM insulating phase for $%
0.48<x\leq {0.50}$.\cite{CEAFM}

In all cases, the 2D character of the magnetic interactions is evident in
these systems. For example, the measured critical exponent of the sublattice
magnetization near $T_{C}$ in the $x=0.40$ sample,\cite{MagIsing} was found
to be very close to the expected value for the 2D Ising model.\cite{Ising}
The two-dimensionality of these systems is also revealed by the strong
anisotropy in the resistivity, with the $ab-$plane resistivity being two or
three orders of magnitude smaller than that along the $c-$axis.\cite%
{BilayersDisc}

We have studied the influence of the dimensionality on the magnetization and
specific heat of the $n=2$ hole-doped bilayered system La$_{2-2x}$Sr$_{1+2x}$Mn%
$_{2}$O$_{7}$ ($0.30\leq {x}\leq {0.50}$). We present a consistent
interpretation of the low-temperature data in terms of a gas of thermally
excited 2D magnons.

\section{Experimental Details}

Single crystals of La$_{2-2x}$Sr$_{1+2x}$Mn$_{2}$O$_{7}$ for $0.30\leq{x}%
\leq {0.50}$ were melt grown in flowing $20\%$ O$_{2}$ (balance Ar) in a
floating zone optical image furnace (NEC SC-M15HD). Magnetization
measurements have been taken in a \textit{Quantum Design} $dc$ SQUID MPMS-5T
magnetometer. The specific heat was measured using the specific heat insert
of a \textit{Quantum Design} PPMS-9T measurement system.

\section{Experimental Results}

\begin{figure}[tbh]
\includegraphics[height=8cm]{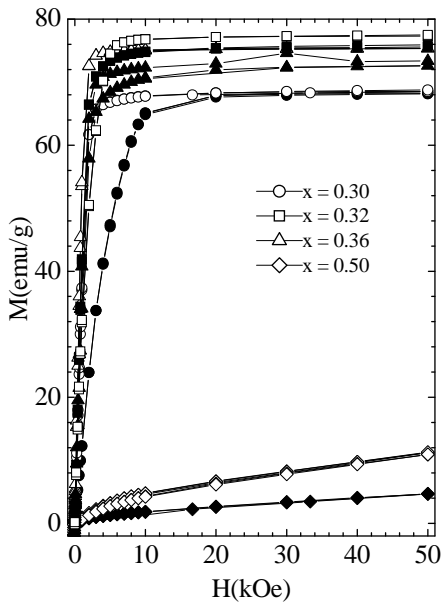}
\caption{$M$vs$H$ measured at $10$K for $x=0.30$, $0.32$, $0.36$ and $0.50$
samples. Solid symbols correspond to $H\parallel$ $ab$ and open symbols to $%
H\parallel$ $c$.}
\label{MvsH}
\end{figure}

Figure \ref{MvsH} shows the field-dependent $dc-$magnetization curves, $M$vs$%
H$, taken at $10$K for $H\parallel $ $ab$ and $H\parallel $ $c$. The
measurements in all samples were made by zero field cooling (ZFC) to $10$%
K, then the field was increased to $5$T and decreased again to zero. For $%
x=0.30,0.32$ and $0.36$ we observed the usual non-linear FM behavior. The $%
x=0.50$ sample displays a linear AFM behavior with a slight anisotropy
between the $c-$axis and the $ab-$plane plus a small non-linear FM-like
behavior at low fields ($H<10$kOe).

From the $M$vs$H$ curves we obtain the saturation magnetization, $M_{S}$,
for the FM samples. Using this values it is possible to estimate the
effective spin of the Mn ions, since $S_{eff}=M_{S}/2\mu _{B}$, and compare
it with the theoretical expected one, $S_{eff}^{theo}$.

Considering that the theoretical valance of the manganese ion in the undoped
compound $x=0$, La$_{2}$SrMn$_{2}$O$_{7}$, is $3+$, and that their
electronic configuration is 3d$^{4}$4s$^{0}$, their expected average
effective spin is $S_{eff}^{theo}(x=0)=2$. Under hole-doping this value will
be reduced by $x/2$, giving $S_{eff}^{theo}(x)=(4-x)/2$. On Table I, we note
that the agreement between $S_{eff}$ and $S_{eff}^{theo}$ is at best
reasonable. That indicates that the average spin model does not provide a
very good description of the saturation magnetization.

\begin{table}[thb]
\caption{$M_{S}$, $S_{eff}$ and $S_{eff}^{theo}$ for the FM samples.}%
\begin{ruledtabular}
\begin{tabular}{cccc}
$x$&$M_{S}$(emu/g)&$S_{eff}$&$S_{eff}^{theo}$ \\
\hline
  0.30 & 68.5 & 1.63 &1.85\\
  0.32 & 77.3 & 1.92 &1.84\\
  0.36 & 75.3 & 1.86 &1.82\\
\end{tabular}
\end{ruledtabular}
\end{table}

\begin{figure}[tbh]
\includegraphics[height=8cm]{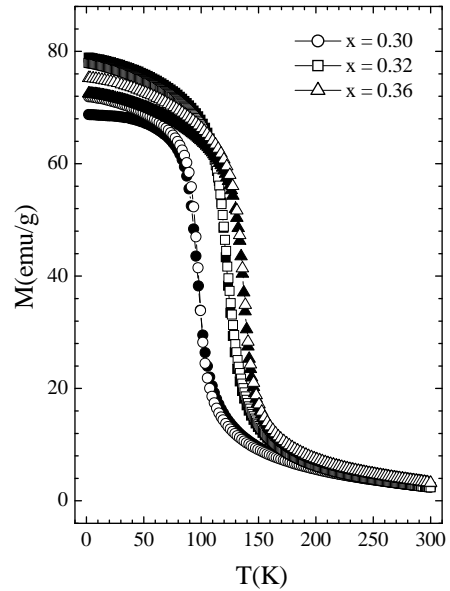}
\caption{$M$vs$T$ for the FM samples $x=0.30$, $0.32$ and $0.36$ taken at $%
H=1$T. The solid symbols correspond to $H\parallel$ $ab$ and open symbols to
$H\parallel$ $c$.}
\label{FMMvsT}
\end{figure}

Figure \ref{FMMvsT} shows the temperature dependence of magnetization, $M$vs$%
T$ curves, for the FM samples. All samples were field cooled from $300$ to $%
1.8$K in a $1$T applied field. The curves show a FM transition at $T_{C}\sim
{100-140}$K, as reported previously,\cite{BilayerRev} with a slight
anisotropy between $H\parallel $ $c$ and $H\parallel $ $ab$, which
approaches zero for $x=0.32$.

\begin{figure}[tbh]
\includegraphics[height=8cm]{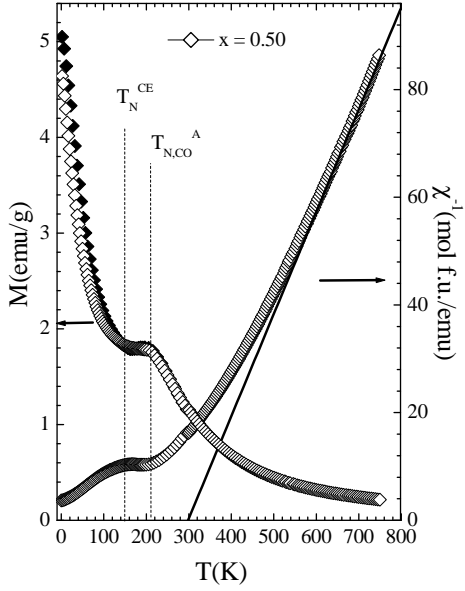}
\caption{\textit{Left scale}: $M$vs$T$ for the AFM sample $x=0.50$ measured
at $H=1$T. The vertical dotted lines represent the N\`{e}el temperatures of
the two AFM phases at $T_{N}^{CE}=145$K for the CE type AFM structure and $%
T_{N,CO}^{A}=210$K for the A type one; \textit{Right scale}: $\protect\chi%
^{-1}$vs$T$ for the AFM sample $x=0.50$, with the solid line showing the
Curie-Weiss fitting for the parameters $p_{eff}= 4.59\protect\mu_{B}/$Mn and
$\protect\theta=300$K. The solid symbols correspond to $H\parallel$ $ab$ and
the open symbols to $H\parallel$ $c$.}
\label{AFMMvsT}
\end{figure}

In the left scale of the figure \ref{AFMMvsT} the $M$vs$T$ curves
for the AFM sample are shown. The sample was field cooled from
$750$ to $1.8$K in a $1$T applied field. As pointed out by Kubota
\textit{et al.}\cite{CEAFM} when analyzing their magnetic neutron
diffraction data, this sample presents two kinds of AFM phases,
an A-type and a CE-type. The N\`{e}el temperature of
the A-AFM phase coincides with the charge ordering temperature at $%
T_{N,CO}^{A}=210$K, while the CE-AFM orders at $T_{N}^{CE}=145$K.\cite{CEAFM}
These two characteristic temperatures are shown in figure \ref{AFMMvsT} by
vertical dotted lines.

The right scale of figure \ref{AFMMvsT} shows the temperature dependence of
the inverse of magnetic susceptibility, $\chi ^{-1}$, for the $x=0.50$
sample for both directions, $H\parallel $ $c$ and $H\parallel $ $ab$. Above $%
T\simeq {100}$K the two curves are identical. The number of
effective Bohr magnetons, $p_{eff}$, and the Curie-Weiss
temperature, $\theta $, extracted from fitting the data to $\chi
^{-1}=(T-\theta )/C$, for the range $600<T<750$K are $p_{eff}=\sqrt{8C}%
=4.59\mu _{B}/$Mn and $\theta =300$K. For this compound, the expected
valance of Mn is 3.5 and $p_{eff}=4.4\mu _{B}$, in good agreement with our
experimental results.

\begin{figure}[tbh]
\includegraphics[height=8cm]{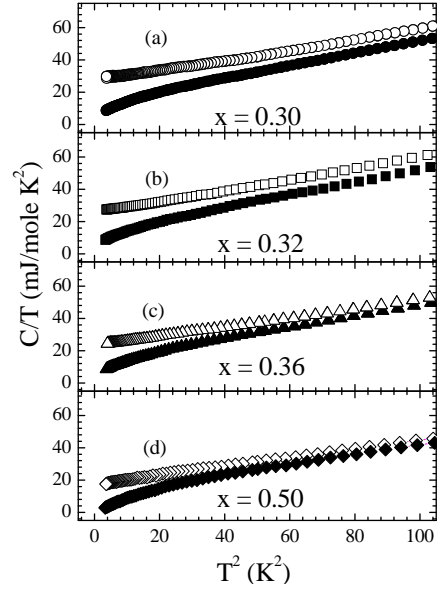}
\caption{$C/T$vs$T^{2}$ for $x=0.30$ (a), $0.32$ (b), $0.36$ (c) and $0.50$
(d) samples. Open symbols represent the zero field measurements. Solid
symbols corresponds to measurements with $H=9$T $\parallel$ $c$.}
\label{Cpcurves}
\end{figure}

Figure \ref{Cpcurves} shows the specific heat results plotted as $C/T$vs$%
T^{2}$, at zero field and $H=9$T, applied parallel to the
$c-$axis for all samples. Two interesting features can be
observed. The first one is the linear dependence of the
zero-field data that shows a high value for gamma. This feature
is basically independent of the hole concentration, magnetic
order and conductivity. The second feature is the strong magnetic
field dependence of the data. In the next section we will show
that the origin of this behavior results from the 2D character of
the system, namely, the influence of the 2D magnons. We will
analyze the influence of the 2D magnons on the magnetization of
the FM samples and on the specific heat of all compounds.

\section{Theoretical Model}

\subsection{Magnetization}

To analyze the change on the magnetic response due to elementary
excitations, it is useful to consider the fractional change of the
magnetization, $\lbrack M(T\rightarrow {0})-M(T)\rbrack /M(T\rightarrow {0}%
)\equiv {\Delta M/M(0)}$. Using boson statistics, it can be
expressed as:\cite{SSBook}

\begin{equation}
\frac{\Delta M}{M(0)}=\frac{S-\langle {S_{z}}\rangle }{S}=\frac{1}{4\pi S}
\int_{0}^{\infty }\frac{kdk}{e^{\frac{\hbar \omega (\mathbf{k})}{k_{B}T}}-1},
\label{MagnonMag1}
\end{equation}
for spin waves in a system with both an acoustic and an optical branch, with
the latter not thermally populated. Here $S$ is the spin, $\langle {S_{z}}%
\rangle $ is the average value of the $z$ component of the spin, $\mathbf{k}$
is the magnon wave vector and $\hbar \omega (\mathbf{k})$ is the magnon
energy.

Considering 2D FM acoustic magnons, with wave vector
$\mathbf{k}=\mathbf{k}_{a}+\mathbf{k}_{b}$ (measured in units of
the reciprocal of the nearest-neighbor distance, $3.87$\AA ), the
dispersion relation at low temperatures, neglecting the
interbilayer exchange energy $J^{\prime }$, in the presence of a
field, can be written as

\begin{equation}
\hbar \omega (\mathbf{k})=-J_{\parallel}S\mathbf{k}^{2}+2\mu
_{B}H=D\mathbf{ k}^{2}+\Delta ,  \label{2DDispRel}
\end{equation}
with $D\equiv {-J_{\parallel}S}$ being the spin wave stiffness
constant and $\Delta \equiv {2\mu _{B}H}$ the gap induced by the
magnetic field.
Evaluating (\ref{MagnonMag1}) using the dispersion relation (\ref%
{2DDispRel}), we obtain

\begin{equation}
\frac{\Delta M}{M(0)}=-\frac{T}{8\pi DS}ln(1-e^{-\frac{\Delta
}{T}}), \label{MagnonMag2}
\end{equation}
with $\Delta $ in units of K.

The corresponding analysis for A-AFM magnons is more complicated and will
not be discussed here.

\subsection{Specific Heat}

The contribution to the specific heat from the magnon excitations is
calculated summing over harmonic oscillator-like contributions $\sim {(\frac{%
\hbar \omega }{ k_{B}T})^{2}\frac{e^{\frac{\hbar \omega }{k_{B}T}}}{(e^{%
\frac{\hbar \omega }{k_{B}T}}-1)^{2}}}$. We can express the total
contribution to the specific heat coming from the acoustic branch\ of the 2D
FM magnons having the dispersion relation (\ref{2DDispRel}) as

\begin{equation}  \label{Cp2D}
C_{mag}=\frac{1}{4\pi k_{B}T^{2}}\int_{0}^{\infty}\frac{k[\hbar%
\omega(k)]^{2}e^{\frac{\hbar\omega(k)}{k_{B}T}}}{[e^{\frac{\hbar\omega(k)}{%
k_{B}T}}-1]^{2}}dk
\end{equation}

Using this expression we can derive the zero-field, $\Delta =0$, 2D FM
magnon specific heat as

\begin{equation}
C_{mag}(H=0)=\frac{\pi Rk_{B}}{24D}T\equiv {\gamma _{mag}T},
\label{Cpmag}
\end{equation}
where $\gamma _{mag}=\pi Rk_{B}/24D$. This expression was first
derived by Colpa\cite{2DFMMagnons} for a 2D Bravais lattice of
spins. The field-dependent specific heat is obtained solving
(\ref{Cp2D}) with $\Delta \neq {0}$. Unfortunately, in this case,
the result is not analytical.

Neutron scattering results by Hirota \textit{et al.}\cite{A-AFM}
indicate that the coupling between the two MnO$_{2}$ sheets,
$J_{\perp }$, is nearly zero for $x=0.50$. Thus, in the analysis
of the specific heat for $x=0.50$, we neglect $J_{\perp }$ so
that the magnetic contribution to the specific
heat is the result of 2D excitations that are effectively FM for energies $%
\sim {k_{B}T}$. We can then use the expression (\ref{Cpmag})
keeping in mind that in this case $\gamma _{mag}=\pi Rk_{B}/12D$,
since there are two degenerate magnon branches for each $k$-value.

Considering the contributions to the specific heat coming from electrons,
phonons, and magnons, the total zero-field specific heat can be described as

\begin{equation}
\frac{C}{T}(H=0)=\gamma _{e}+\gamma _{mag}+\beta T^{2}\equiv {\gamma
_{eff}+\beta T^{2}},  \label{Cp0TFM}
\end{equation}
where we have defined an effective gamma-term $\gamma _{eff}=\gamma
_{e}+\gamma _{mag}$. So, the zero field specific heat can be displayed as a
linear plot in a $C/T$vs$T$ $^{2}$ curve.

For $H\neq{0}$ the total specific heat will be

\begin{equation}
\frac{C}{T}(H\neq{0})=\gamma_{e}+\beta T^{2}+\frac{1}{4\pi k_{B}T^{3}}%
\int_{0}^{\infty}\frac{k[\hbar\omega(k)]^{2}e^{\frac{\hbar\omega(k)}{k_{B}T}}%
}{[e^{\frac{\hbar\omega(k)}{k_{B}T}}-1]^{2}}dk  \label{Cp9T}
\end{equation}
with $\hbar\omega(k)=Dk^{2}+\Delta$.

For $x=0.50$, the prefactor in front of the integral in eq.(\ref{Cp9T}) need
to be multiplied by a factor of $\ 2$ to account for the two-fold degeneracy
mentioned previously.

In the next section we will apply these results to the analysis
of the magnetization and specific heat data.

\begin{figure}[t]
\includegraphics[height=8cm]{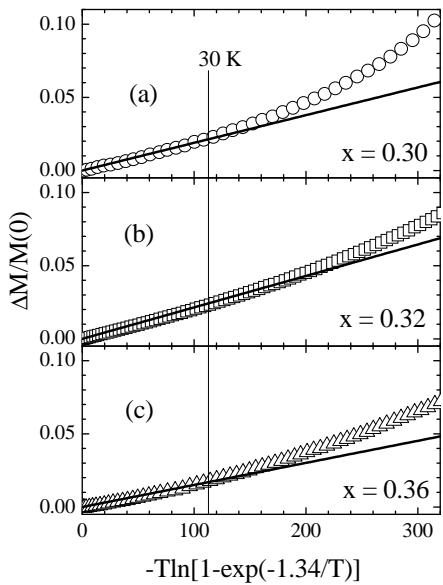}
\caption{$\Delta M/M(0)$ plotted as function of $-Tln[1-e^{-\frac{\Delta }{T}%
}]$ for $x=0.30$ (a), $0.32$ (b), and $0.36$ (c). The magnetic field of $H=1$%
T ($\Delta =1.34$K) was applied along the easy-axis. The vertical dotted
line indicates the temperature of $30$K. The solid line represents the
fitting of $\Delta M/M(0)$ to $-bTln[1-e^{-\frac{\Delta }{T}}].$}
\label{DeltaM}
\end{figure}

\begin{figure}[b]
\includegraphics[height=8cm]{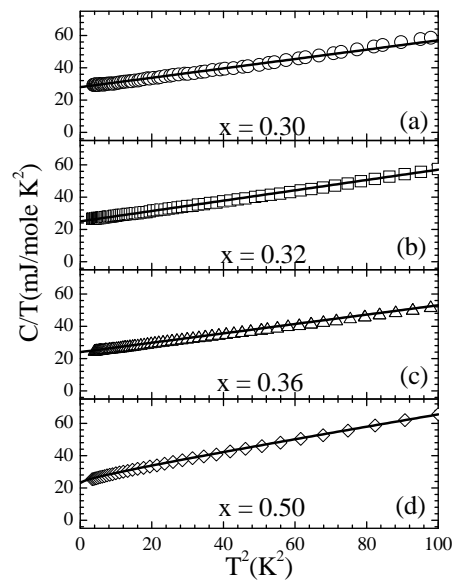}
\caption{Zero field specific heat data for $x=0.30$ (a), $0.32$ (b), $0.36$
(c) and $x=0.50$ (d) samples. The solid lines represents the fitting of the
data to eq. (6).}
\label{CpFit}
\end{figure}

\section{Analysis and Discussion}

\subsection{Magnetization}

Figure \ref{DeltaM} shows the fractional change of the
magnetization $\Delta M/M(0)$ as function of
$-Tln[1-e^{-\frac{\Delta }{T}}]$ for the FM samples. A field of
$H=1$T ($\Delta =1.34$K) was applied parallel to the easy-axis of
the samples: $\parallel $ $c$ for $x=0.30$ and $\parallel $ $ab$
for $x=0.32$ and $0.36$. We fit the data to $\Delta
M/M(0)=-bTln[1-e^{-\frac{\Delta }{T}}]$ and the parameters
obtained are given in Table II.

\begin{table}[tbh]
\caption{Parameters $b$, $D$ and $J_{\parallel }$, obtained as
described in the text.}
\begin{ruledtabular}
\begin{tabular}{ccccc}
$x$&$b(10^{-4}K^{-1})$&$D(K)$&$J_{\parallel}(K)$\\
\hline
  0.30 & 1.9(1)& 112(6) &-60(2)\\
  0.32 & 2.1(1) & 104(6) &-56(2)\\
  0.36 & 1.6(1) & 138(9) &-76(4)\\
\end{tabular}
\end{ruledtabular}
\end{table}

From the expression (\ref{MagnonMag2}) we have that $b=1/8\pi DS$.
In order to enable us a consistent comparison between our model
and the neutron diffraction analysis from literature,\cite
{A-AFM,Chatterji1,Chatterji2,Perring} we use the average
theoretical effective spin $S_{eff}^{theo}$, instead of the
experimental value $S_{eff}$. Using $S_{eff}^{theo}$ we obtained
the stiffness constant,$D$ , and the exchange energy
$J_{\parallel }$ given in Table II. No significant differences
are found if $S_{eff}$ is used instead.
In that case the values of $J_{\parallel }$ are: $-78(4)$, $-52(2)$ and $%
-74(4)$K, for $x=$$0.30$, $0.32$ and $0.36$, respectively. Our values of $%
J_{\parallel }$ are in good agreement with the $J_{\parallel }$
values obtained from neutron diffraction measurements\cite
{A-AFM,Chatterji1,Chatterji2,Perring}. Thus, from our analysis we
can conclude that the main contribution to the temperature
dependence of the magnetization of the FM bilayer system comes
from the 2D FM magnons.

\subsection{Specific heat}

Figure \ref{CpFit} shows the zero field specific heat data fit to eq.(\ref%
{Cp0TFM}). In the temperature interval between $1.8$K$\leq T\leq 10$K, the
data are in a good agreement with the FM magnon model. From this analysis we
derived the $\gamma _{eff}$ and $\beta $ values for all concentrations.

\begin{table}[hbt]
\caption{Experimental specific heat parameters.}%
\begin{ruledtabular}
\begin{tabular}{ccccc}
$x$& $\gamma_{eff}$ & $\gamma_{e}$ & $\beta$ & $D(H=9$T$)$\\
& (mJ/mole K$^{2}$) & (mJ/mole K$^{2}$)& (mJ/mole K$^{4}$)& (K)\\
\hline
  0.30 & 28(1) & 7.0(2) & 0.27(3) & 88(4)\\
  0.32 & 25(1) & 7.0(3) & 0.31(3) & 96(2)\\
  0.36 & 23(1) & 7.0(2) & 0.27(2) & 98(6)\\
  0.50 & 24(1) & 4.0(8) & 0.40(4) &196(16)\\
\end{tabular}
\end{ruledtabular}
\end{table}

\begin{table*}[tbh]
\caption{ Parameters obtained from the specific heat analysis as described
in the text.}%
\begin{ruledtabular}
\begin{tabular}{cccccc}
$x$ & $g(\varepsilon_{F})$ & $\gamma_{mag}$ & $D(H=0)$ & $J_{\parallel}(H=0)$ & $J_{\parallel}(H=9$T$)$ \\
& (states/ Ry f.u.) & (mJ/mole K$^{2}$) & (K) & (K) & (K)\\
\hline
  0.30 & 40(1) & 21(1) & 104(2)& -56(2) & -47(1)\\
  0.32 & 40(2) & 18(1) & 122(2)&  -66(4) & -52(2)\\
  0.36 & 40(1) & 16(1) & 136(2)&  -74(4) & -54(2)\\
  0.50 & 23(5) & 20(1) & 218(4)& -124(6)& -112(8)\\
\end{tabular}
\end{ruledtabular}
\end{table*}

\begin{figure}[t]
\includegraphics[height=8cm]{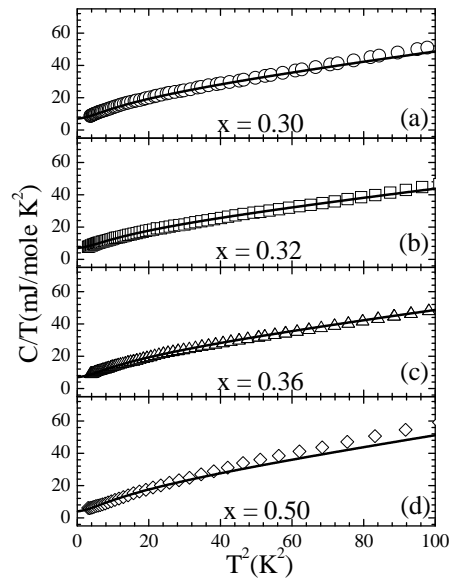}
\caption{Fit of the specific heat data measured at $H=9$T $\parallel$ $c$ to
eq.(7) for $x=0.30$ (a), $0.32$ (b), $0.36$ (c), and $0.50$ (d) samples.}
\label{9TCpfit}
\end{figure}

Figure \ref{9TCpfit} shows the specific heat data, taken at
$H=9$T, fitted to eq.(\ref{Cp9T}). In this fitting we used the
$\beta $ parameters derived from the zero field analysis. From it
we obtain $\gamma _{e}$ and $D(H=9$T$)$ and calculated the
exchange energy $J_{\parallel }(H=9$T$)$. No appreciable
difference is found in the $J_{\parallel }$ values when using $%
S_{eff}^{theo} $ or $S_{eff}$. For $x=0.50$, it should be noted that the
ferromagnetic approximation for the spin wave energies, as indicated by the
linear behavior of the specific heat in zero field, only pertains to the
modes that make a significant contribution in the temperature range of the
experiment. If measurements were made at much lower temperatures, one would
expect a temperature dependence characteristic of a weakly anisotropic A-AFM.

We calculated the $\gamma _{mag}$ from $\gamma _{mag}=\gamma _{eff}-\gamma
_{e}$. Using $\gamma _{e}$ and $\gamma _{mag} $ we computed the electron
density of states at Fermi level, $g(\varepsilon _{F})$, and $D(H=0)$.
Finally, from $D(H=0)$ we obtain the exchange energy $J_{\parallel}(H=0)$.
The experimental parameters $\gamma_{eff}$, $\gamma_{e}$, $\beta$ and $D(H=9$%
T$)$ are listed in the Table III. The values of $g(\varepsilon_{F})$, $%
\gamma_{mag}$, $D(H=0)$, $J_{\parallel}(H=0)$ and
$J_{\parallel}(H=9$T$)$ derived from the data, are given in the
Table IV.

The values of $\gamma _{e}$ and $g(\varepsilon _{F})$ of the insulating
samples, $x=0.30,$ $0.32$ and $0.36$ are in reasonable agreement with those
reported in the literature (see, e.g., the results of Okuda \textit{et al.}%
\cite{Okuda}), suggesting a half metallic behavior. In fact, electronic band
structure calculations performed by Meskine \textit{et al.}\cite{Band}
suggested a half metallic character for some bilayers. In particular, for $%
x=0.50$ they found the theoretical value of $g(\varepsilon _{F})=19$
states/Ry f.u., in excellent agreement with our experimental value of $23$
states/Ry f.u..

\begin{figure}[bth]
\includegraphics[height=8cm]{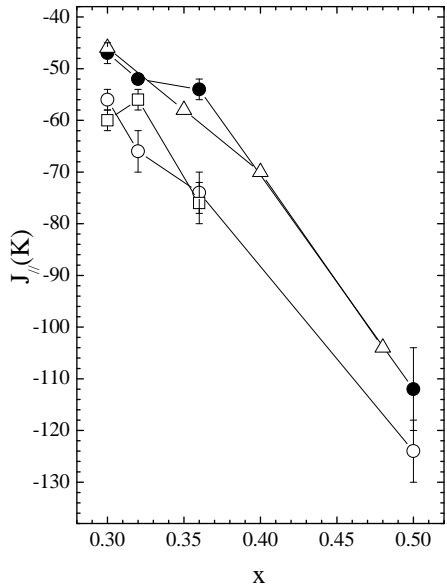}
\caption{Comparison between the $J_{\parallel}$ values, obtained
on the present work and the ones reported by Hirota \textit{et
al}.\cite {A-AFM} Open squares correspond to the values derived
from the $MvsT$
analysis. Open (solid) circles correspond to the values derived from $%
C/TvsT^{2}$ analysis with $H=0(9$T$)$. Open triangles present the
values from Hirota \textit{et al}.\cite{A-AFM}} \label{comparison}
\end{figure}

The values of $J_{\parallel }(H=0)$ and $J_{\parallel }(H=9$T$)$
are in reasonable agreement with the results previously reported
by neutron scattering
studies\cite{A-AFM,Chatterji1,Chatterji2,Perring}. In figure
\ref{comparison} we compare the $J_{\parallel }$ values obtained
in the present work and the ones reported by Hirota \textit{et
al.}\cite{A-AFM} The $J_{\parallel }(H=0)$ values obtained from
the zero field specific heat agree with the values inferred from
the analysis of the magnetization. The values of $J_{\parallel
}(H=9$T$)$ differ somewhat from $J_{\parallel }(H=0)$. That
may be a consequence of the presence of the double exchange mechanism since $%
J_{\parallel }$ is field-independent for a pure Heisenberg system. The $%
J_{\parallel }(H=0)$ values are also in fair agreement with the Hirota\'{ }s
ones, while the $J_{\parallel }(H=9$T$)$ are in excellent agreement. The
above is possible related to the differences in the energy scale probed by
the different techniques. Our zero field thermodynamic measurements probe
lower energy magnons, while the field-dependent measurements involve magnons
with a slight higher energy, as the $9$T magnetic field opens a gap $\sim 1$%
meV in the magnon dispersion relation. On the other hand, generally neutron
experiments involve magnons with energies $\gtrsim 3$meV. So, it is expected
a smaller difference between the values of $J_{\parallel }$ derived from
neutron diffraction and the obtained from $9$T specific heat than with the
ones extracted from the zero field specific heat. Also, a significant
renormalization of the magnon energies at small $k$, that is not reflected
in the neutron data, may be present.

\section{Conclusion}

In summary, our experimental specific heat results reveal two interesting
features: (i) the linear, high-gamma behavior of the zero-field data,
independent of hole concentration, magnetic order or conductivity and (ii)
the strong magnetic field dependence of the measurements.

We propose a model considering the presence of 2D FM magnons to explain the
data. We calculated the influence of these magnons on the magnetization of
the FM samples and specific heat. In all the cases the model gives a good
fit, showing that the change in magnetization and the specific heat of the
system can be well described by a 2D FM magnon gas. In the case of the
antiferromagnetic system ($x=0.50$), the linear behavior of the specific
heat is shown to be a consequence of the weak coupling between the bilayers
sheets, that enables the occurrence of the 2D FM magnons.

As noted, we find similar values for $J_{\parallel}$ from both the
magnetization and the specific heat analysis, indicating an internal
consistency on our analysis. Moreover, the values of $J_{\parallel}$ we
found are close to the ones derived from the neutron scattering studies\cite%
{A-AFM,Chatterji1,Chatterji2,Perring}. The last provides further support to
our analysis.

\section{Acknowledgments}

This work was supported by the Brazilian Agencies CNPq and FAPESP and NSF
01-02235. D. L. Huber would like to thank T. G. Perring and N. Shannon for
helpful comments.

\end{document}